 \documentclass[twocolumn]{ceurart}
\usepackage{listings}
\begin{document}

\copyrightyear{2023}

\copyrightclause{Copyright for this paper by its authors. Use permitted under Creative Commons License Attribution 4.0 International (CC BY 4.0).}

\conference{Joint Workshops at 49th International Conference on Very Large Data Bases (VLDBW’23) — Second International Workshop on Composable Data Management Systems (CDMS’23), August 28 - September 1, 2023, Vancouver, Canada}

\title{Building a serverless Data Lakehouse from spare parts}

\tnotemark[1]
\tnotetext[1]{With our title we pay our tribute to the paper by the Firebolt team \cite{DBLP:conf/vldb/PasumanskyW22}. While our goals, timeline and methodology are different, our work shares the underlying philosophy.}


\author[1,2]{Jacopo Tagliabue}[%
orcid=0000-0001-8634-6122,
email=jacopo.tagliabue@nyu.edu
]
\cormark[1]

\address[1]{Bauplan, New York City, United States}
\address[2]{Tandon School of Engineering, NYU, New York City, United States}

\author[1]{Ciro Greco}[%
orcid=0009-0007-0359-4130,
email=ciro.greco@bauplanlabs.com
]

\author[1]{Luca Bigon}[%
orcid=0009-0001-0028-7983,
email=luca.bigon@bauplanlabs.com
]
\fnmark[1]

\cortext[1]{Corresponding author.}

\begin{abstract}
    The recently proposed Data Lakehouse architecture is built on open file formats, performance, and first-class support for data transformation, BI and data science: while the vision stresses the importance of lowering the barrier for data work, existing implementations often struggle to live up to user expectations. At \textit{Bauplan}, we decided to build a new serverless platform to fulfill the Lakehouse vision. Since building from scratch is a challenge unfit for a startup, we started by re-using (sometimes unconventionally) existing projects, and then investing in improving the areas that would give us the highest marginal gains for the developer experience. In this work, we review user experience, high-level architecture and tooling decisions, and conclude by sharing plans for future development.
\end{abstract}


\maketitle

\section{Introduction}

\cite{Zaharia2021LakehouseAN} argues that the popular data warehouse architecture will soon be replaced by a new architectural pattern, the Data Lakehouse (DLH). A DLH is built on open file formats (e.g. Parquet), exceptional performance, and first-class support for engineering (data transformation), analytics (BI) and inferential (data science) use cases. The vision of such architecture is first and foremost about flexibility, making it possible for organizations to choose different ways to operationalize data depending on data volumes, use cases, and technological and security constraints. 

This is particularly valuable for large organizations where data democratization is crucial to achieve agility \cite{Dehghani2022}: enabling easier access to and understanding of data is the prerequisite for organizations to best leverage their data. The heterogeneity of use cases is reflected in the complexity of the underlying infrastructure (Fig. \ref{fig:lakehouse}), with some pieces coming from databases (query engines, tables, data catalogs etc.), some from distributed systems (pipeline, orchestration, runtime management and optimization etc.).

There are two primary approaches to realize the DLH vision. The first is improving the usability and flexibility of existing Big Data technologies: e.g., one could start by adding automated cluster configurations to Apache Spark. Although everyone will stand behind easier development in Spark, this approach falls short of delivering a developer experience truly aligned with the vision of the DLH, as we will discuss further below.

A different approach would consist in building a system from scratch based on foundational principles, while maintaining storage as a separate component; e.g., one could imagine dispensing with the Java Virtual Machine (JVM) altogether, under the assumption that the advantages of using it are not significant enough, if no legacy is involved. This approach is unfortunately impractical to the point of being unattainable: re-building a catalog, a query engine and multi-language runtimes at the same time is an unreasonable amount of work for a resource-constrained startup.

In \textit{this} paper we describe how we designed \texttt{Bauplan}, a serverless platform implementing the DLH vision by putting the development experience first. We built the first version of \texttt{Bauplan} by following a third approach, i.e. re-purposing (sometimes unconventionally) existing tools when possible, and investing most resources into differentiating features. We challenge the proponents of the first approach, arguing that the data landscape has undergone significant changes since the Big Data era and a new foundation is needed. Similarly, we object that a complete rewrite is unnecessary. Instead, we argue that a well-informed industry perspective can effectively narrow down the problem scope and lay a new foundation beyond traditional Big Data frameworks. 

\section{A Practitioner Perspective}

Aside from security and compliance, the biggest argument in favor of the DLH is flexibility: different teams can use different tools to process data for different use cases. Practically, this implies that any DLH needs to support two different use cases:

\begin{itemize}
    \item\textbf{Query and Wrangle} (QW), referring to the scenario where users need to explore data and ask specific questions (e.g. counting how many marketing emails were opened in the previous month). Querying predominantly involves SQL, while Wrangling is often performed in Python.
    \item\textbf{Transform and Deploy} (TD), referring to the scenario where users need to construct code-driven, reproducible data pipelines (DAGs) that generate new artifacts for downstream utilization. For instance, building a dashboard exposing the performances of marketing emails across different user demographics. Due to the distinct strengths and weaknesses of SQL and Python, the combination of both is often optimal.
\end{itemize}

Importantly, depending on the phase in which developers find themselves in the development cycle, their way to interact with the data can be either Synchronous or Asynchronous. While QW is \textit{de facto} always synchronous, TD tend to be more nuanced and need to support both.

\begin{itemize}
    \item \textbf{Synchronous} is when a user issues a command (a SQL query, or a DAG run) and awaits for the results to come back. In this scenario, simplicity and fast feedback loop are the key goals \cite{https://doi.org/10.48550/arxiv.2209.09125};

    \item \textbf{Asynchronous} is when a command is issued (often by another system, such as an orchestrator) and the user is involved in monitoring the outcome at a later time. In this scenario, reliability, resilience and infrastructure ergonomics are the key goals.
\end{itemize}

\begin{table}[h]
  \caption{Use cases and interaction modalities in the data life cycle: development vs production.}
  \label{tab:taxonomy}
  \begin{tabular}{lcl}
    \toprule
    Use case & Env & Mode\\
    \midrule
    Querying + Wrangling  & Dev & Synch \\
    Querying + Wrangling  & Prod & Synch \\
    Transforming + Deploying & Dev & Synch + Asynch \\
    Transforming + Deploying  & Prod & Asynch \\
    \bottomrule
  \end{tabular}
\end{table}

The interplay between use cases and modalities is summarized in Table~\ref{tab:taxonomy}. A DLH needs to provide a coherent developer experience across the different phases of their development cycle (Dev vs. Prod) while supporting the possibility of interacting with data (QW vs. TD) in both synchronous and asynchronous ways.

To achieve this, we designed \texttt{Bauplan} with the following general design principles in mind:

\begin{itemize}
\item \textbf{Serverless experience}: to fully leverage the separation of storage and compute, developers should deal with as little infrastructure as possible. We propose to decouple data logic from execution to enable a ``serverless'' experience based on a declarative approach; furthermore, since data pipelines are functional in nature (output of parent nodes is input for children), a function-as-a-service deployment is \textit{prima facie} a natural fit.\footnote{Note that we purposely use the term with some flexibility (Section \ref{sec:run}).}

\item \textbf{Software development patterns}: it is often the case that the only developers who can bring data applications to production are those who possess a special data engineering skill set. Empowering developers with more general coding skills to do impactful work on data is a fundamental piece of the DLH vision. Systems should allow users to use only familiar tools like SQL, standard Python, the CLI and Git.

\item \textbf{Reproducibility and versioning}: because the primary factor for building data products involves reproducible and versioned code pipelines, we embrace the idea that code provides a (mostly declarative) way to build data. Likewise, data is treated as code, adopting a life cycle including branching, committing, and merging.

\item \textbf{Full Auditability}: cloud clusters with long startup time and complex configurations encourage developers to resort to local development to expedite the feedback loop. However, this pattern exacerbates the challenges of software development (e.g. dependency management) while introducing potential security issues. We advocate for a cloud-first approach, ensuring that all work and access are centralized, auditable, and aligned with security and governance policies.
\end{itemize}

\section{Departing from Spark}

Before delving into the specifics of our design for \texttt{Bauplan}, we wish to explain the rationale behind departing from Spark, which is widely regarded as the industry standard for analytics at scale and holds a significant position in numerous DLH implementations.

Given our discussion about the ideal DLH developer experience, we believe Spark falls short for several structural reasons. For instance, slow startup and execution makes Spark sub-optimal for synchronous operations, such QW. At the same time, the system has a notoriously steep learning curve \cite{ChambersandZaharia18, Damjietal2020}, both from an API and a debugging perspective: when thinking about TD, it is often hard to reason about it \cite{Wang2021UnderstandingTC,9007378}. If the DLH vision is truly about enabling a broader set of practitioners to perform data transformations, these systems are not necessarily the best design choice. 

\subsection{The Reasonable Scale hypothesis}
There is an additional point that further strengthens our argument in favor of fast and efficient implementations. The definition of Big Data changed from the time Spark was first introduced: popular datasets from the Big Data era \cite{doi:10.1089/big.2013.0037} (and even recent deep learning challenges \cite{CoveoSIGIR2021,https://doi.org/10.48550/arxiv.2207.05772}) can now be processed comfortably in one machine. 

Most of what we considered internet scale at the time would be considered a ``reasonable scale'' today. The term ``Reasonable Scale'' (RS) has gained popularity within the ML community to describe development practices that stand in contrast to those required by the scale of Big Tech companies. Because the vast majority of companies typically deal with datasets ranging from a few thousands rows to tens of GB, the RS has been used to forcefully argue against large scale distributed systems \cite{10.1145/3460231.3474604}, promoting a pragmatic approach not dissimilar from the seminal COST paper \cite{McSherry2015ScalabilityBA}.

\begin{figure}
  \centering
  \includegraphics[width=0.48\textwidth]{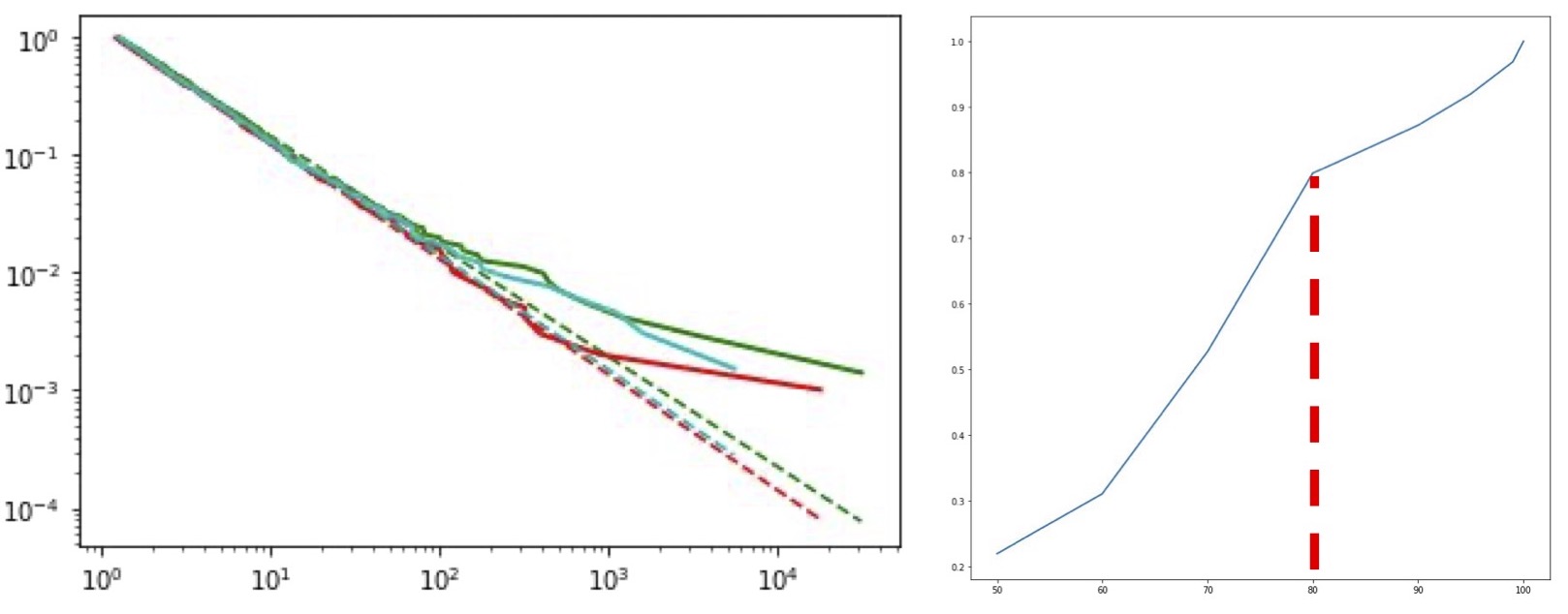}
  \caption{\textit{Left}: a log-log plot for the CCDF (complementary cumulative distribution function) of SQL query times (seconds), from a month in the query history log of three companies (solid lines are the empirical distributions, dotted lines are the fitted versions). \textit{Right}: cumulative cost (y-axis) of running queries up until a given percentile (x-axis): queries up until the 80\textsuperscript{th} percentile (red dotted line) for bytes scanned are responsible for ~80\% of all credit usage.}
  \label{fig:rscale}
\end{figure}

At \texttt{Bauplan}, we investigated the RS hypothesis for data transformation workloads. To accomplish this, we developed scripts that analyzed query patterns and generated reports on query time distributions for SQL workloads in one month. Query time correlates with byte scans and table size, hinting at a power-law distribution where most workloads handle small data volumes. Figure~\ref{fig:rscale} (left) displays distributions for three sample companies, spanning startups to public firms\footnote{To fully anonymize the dataset, we used the \texttt{powerlaw} package \cite{Alstott2013powerlawAP} for distribution fitting: final data are then generated by sampling from the distribution.}. As clear from the log-log plot, the power-law-like behavior holds for all companies, with a good chunk of the queries being run in the $10^0$-$10^1$ seconds range. From one design partner, we were able to also obtain direct estimates about bytes scanned when querying, and associated cloud costs: knowing that the 80\textsuperscript{th} percentile in the bytes distribution corresponds to approximately 750MB, Fig.~\ref{fig:rscale} (\textit{right}) further strengthens the applicability of the RS hypothesis to data workloads.

Finally, the cost of RAM and high-performance disk space keeps decreasing: in the last 10 years, the cost of 1 TB of memory decreased from 5,000 USD to 2,000 USD.\footnote{\url{https://ourworldindata.org/grapher/historical-cost-of-computer-memory-and-storage?time=2010..latest&facet=metric}.} Taken together, these considerations validate RS workloads as an initial viable market for \texttt{Bauplan}, and cast doubt on the necessity of assuming extensive parallel computing is needed for most DLH use cases.

\section{Core Components}

\begin{figure}
  \centering
  \includegraphics[width=0.35\textwidth]{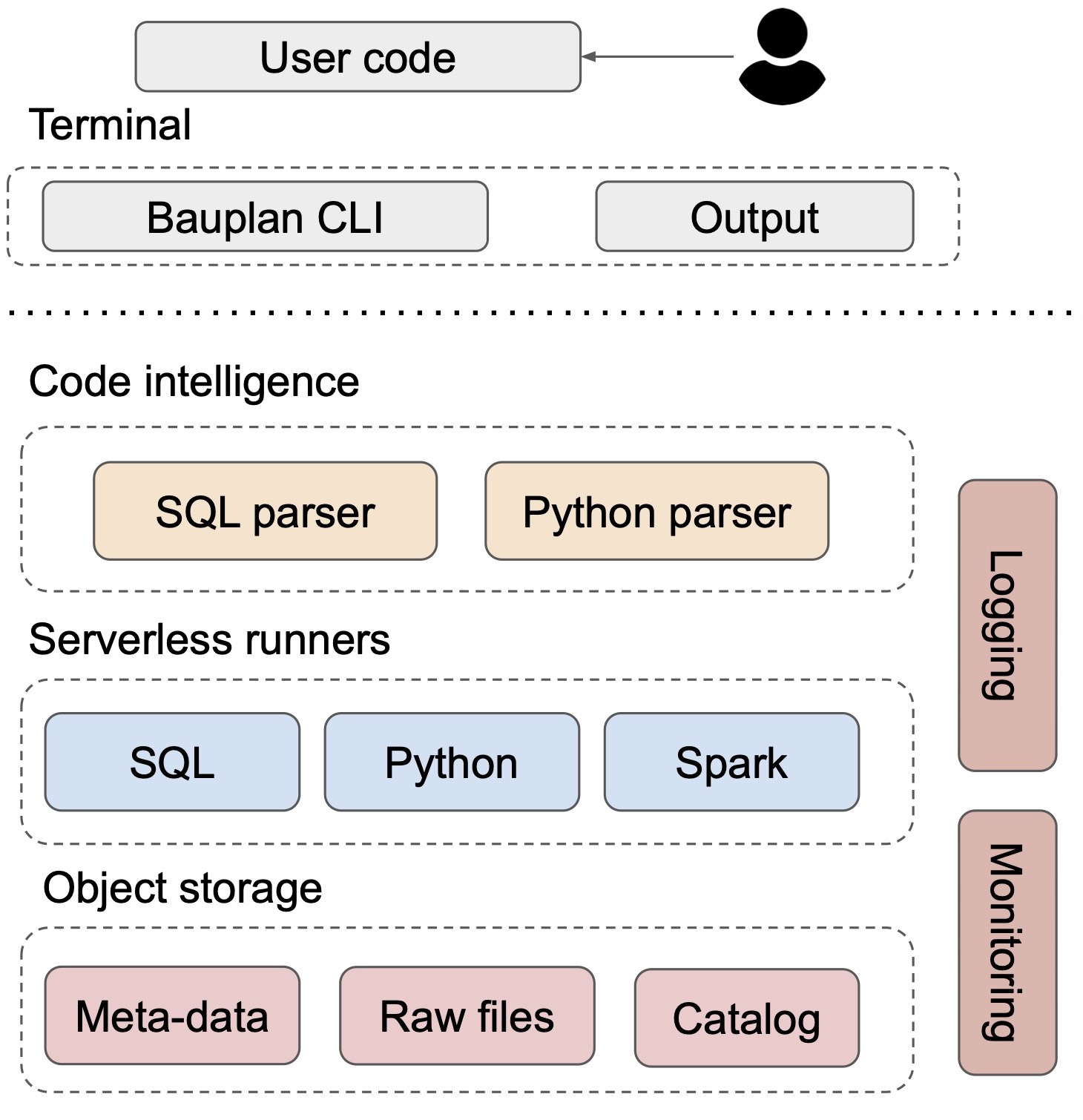}
  \caption{A Bauplan lakehouse and its main components. User's primary interaction mode is code and CLI, while the underlying platform abstracts away pipeline planning, execution and materialization.}
  \label{fig:lakehouse}
\end{figure}

We survey the core components of \texttt{Bauplan} architecture and discuss the rationale behind our choices. The general design is illustrated in Fig.~\ref{fig:lakehouse}: at the top, we have the user layer - including just the code and a CLI; at the bottom, we have the cloud layer, including storage for raw data and derived assets, code intelligence to transform user code into a logical plan and infrastructure to run the actual computations.

\subsection{A sample data pipeline}
\label{sec:example}

\begin{figure*}
  \centering
  \includegraphics[width=0.75\textwidth]{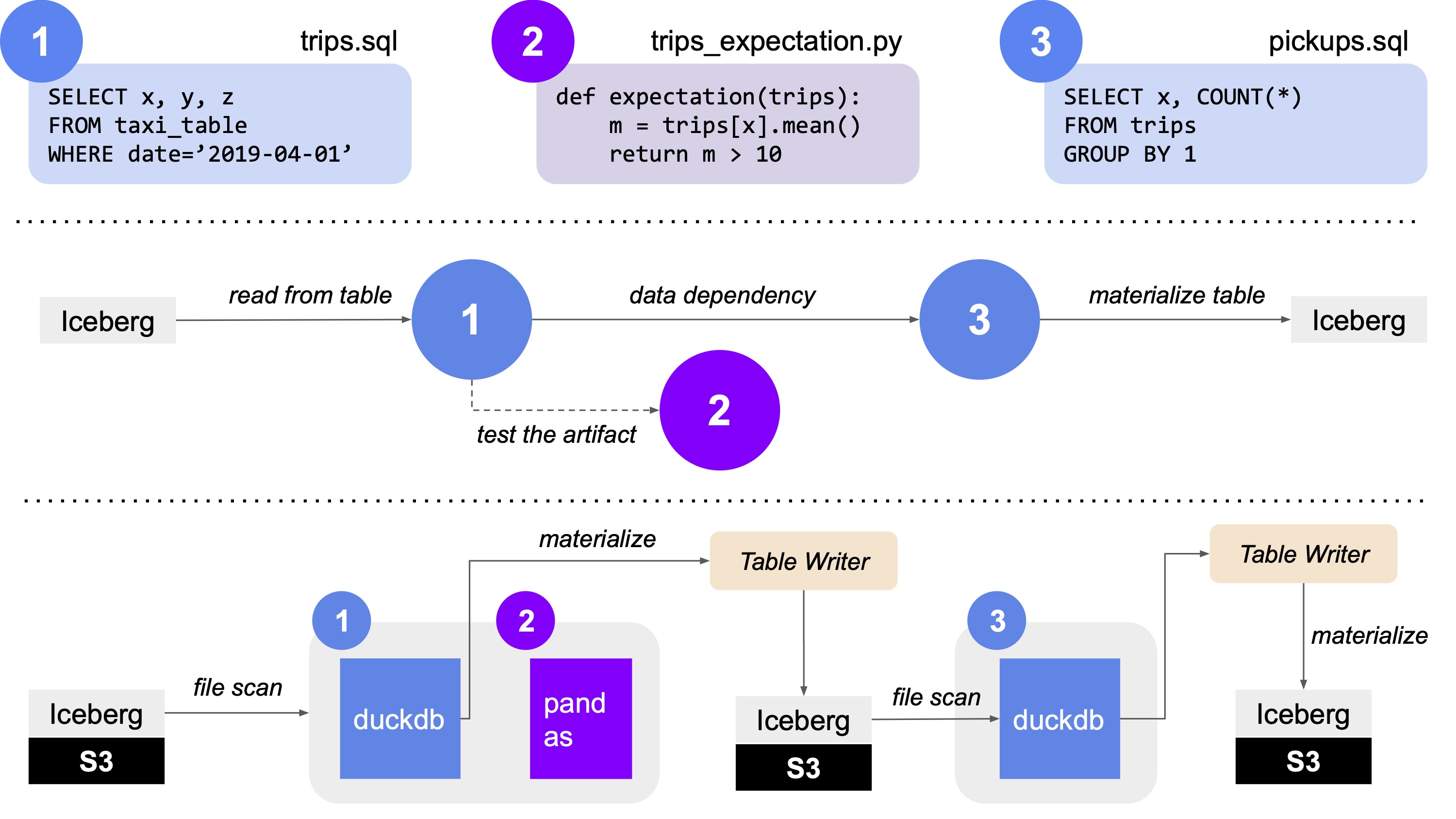}
  \caption{A worked out example of a data pipeline in \texttt{Bauplan}, at three layers of abstraction: \textit{top}, the developer layer, i.e. a modular, multi-language coding pattern in which DAG dependencies are expressed implicitly as part of code dependencies; \textit{middle} the logical plan, i.e. a series of functional operations with explicit dependencies between steps and connection to outside artifacts (e.g. Iceberg tables); \textit{bottom}, the physical plan, i.e. a series of commands (and compatible engines) to satisfy the logical plan in an efficient way - for example, by leveraging data locality, the code in Step 2 can be run without any data movement right after Step 1.}
  \label{fig:plans}
\end{figure*}

Throughout the paper, we use a prototypical data pipeline as a working example. In Fig~\ref{fig:plans} we introduce a small, but functioning data pipeline with two SQL nodes for data artifacts and one Python expectation test, checking the quality of one artifact -- we also report the full SQL and Python code in the \textit{Appendix}.

Without loss of generality, we are simulating a \textit{Transforming} use case over the NYC taxi dataset\footnote{https://www.nyc.gov/site/tlc/about/tlc-trip-record-data.page}: in particular, we transform raw data from an Iceberg table (Section~\ref{sec:catalog}) named \texttt{taxi\_table}, into a final table named \texttt{pickups}, which contains pre-computed popular pickup locations ready to power a dashboard. There are three main notions at play in this DAG:

\begin{itemize}
    \item \textit{the data lake}: while not obvious from the code itself, there is an object storage layer containing the raw data we are starting from: from the developer perspective, users would only interact with logical constructs, such as \texttt{taxi\_table}; from an implementation standpoint, handling persistent object transparently is a huge component of the DLH (Sections \ref{sec:catalog} and \ref{sec:versioning} below).
    
    \item \textit{declarative data assets}: we subscribe to the one-query, one-artifact pattern popularized by dbt-style transformations\footnote{\url{https://github.com/dbt-labs/dbt-core}}: users define artifacts one by one as SQL queries, and the platform builds up the DAG based on parsing and naming convention (Section \ref{sec:intelligence}). Importantly, no imperative-style DAG construction is needed: insofar as users implicitly link together parent and children nodes through their code, functions ``are all you need''\footnote{See also the Appendix for the full code example.};
    
    \item \textit{data expectations}: it is best practice to test the tables produced by a DAG for statistical anomalies. This provides the foundation of the \textit{transform-audit-write} pattern for data development (Section \ref{sec:versioning}): just as in software we can debug, test and even run different versions of an application in parallel against production, automated testing and versioning becomes the foundation of the same approach for data pipelines.\footnote{
    Following the software analogy further, expectations are akin to integration tests, where a new component is embedded in an existing system, and checks are made to ensure the desired output is achieved. A related but different concept is \textit{unit tests}, which instead work on manually fabricated input-output pairs, to test edge cases or important scenarios \textit{irrespective} of the system actually seeing this input. Given our abstractions, Bauplan can easily accommodate both types of tests, especially considering that Python primitives for creating unit tests for tables are better than SQL.}
\end{itemize}

\subsection{Table Format}
\label{sec:catalog}

While a data lake is ultimately made of files, we wish to provide table-like abstractions to our users: by decoupling the actual storage of the data (the file \textit{s3://my-bucket/taxifile.parquet}) from their logical function (the list of taxi trips in NYC), we can reuse the same code across data versions: every command that points to \texttt{taxi\_table} can be executed over different versions of the table with just a configuration change (Section~\ref{sec:versioning}). After considering \textit{Delta Lake}, \textit{Hudi}\footnote{\url{https://hudi.apache.org/}}, and \textit{Iceberg}\footnote{\url{https://github.com/apache/iceberg}} as possible formats to give table-like semantics to the object storage, we chose Iceberg mainly for three reasons: larger community support, full support for time-travel and versioning semantics, limited but increasing compatibility with Python\footnote{\url{https://py.iceberg.apache.org/}}.

At the time of writing, major formats have full read / write support only for JVM engines (e.g. Spark, Presto, Dremio). Considering our working hypothesis about the Reasonable Scale and \texttt{Bauplan} focus on a serverless experience, two major tasks had to be completed to overcome these constraints. First, when running a query over an Iceberg table, our code intelligence module needs to first parse SQL into a table scan to obtain a dataframe-like object (Section~\ref{sec:execution}); second, when materializing a data asset from the DAG back to the data catalog, a Spark session is created to handle the Iceberg \texttt{INSERT}: following our \textit{no infrastructure} principle, we created custom containers (Section~\ref{sec:run}) optimized for starting a Spark command with 300 milliseconds latency -- as a result, the materialization step looks no slower than running any other Python function (as opposed to waiting for a Spark cluster to launch).

\subsection{Data catalog and versioning}
\label{sec:versioning}

\begin{figure}
  \centering
  \includegraphics[width=0.25\textwidth]{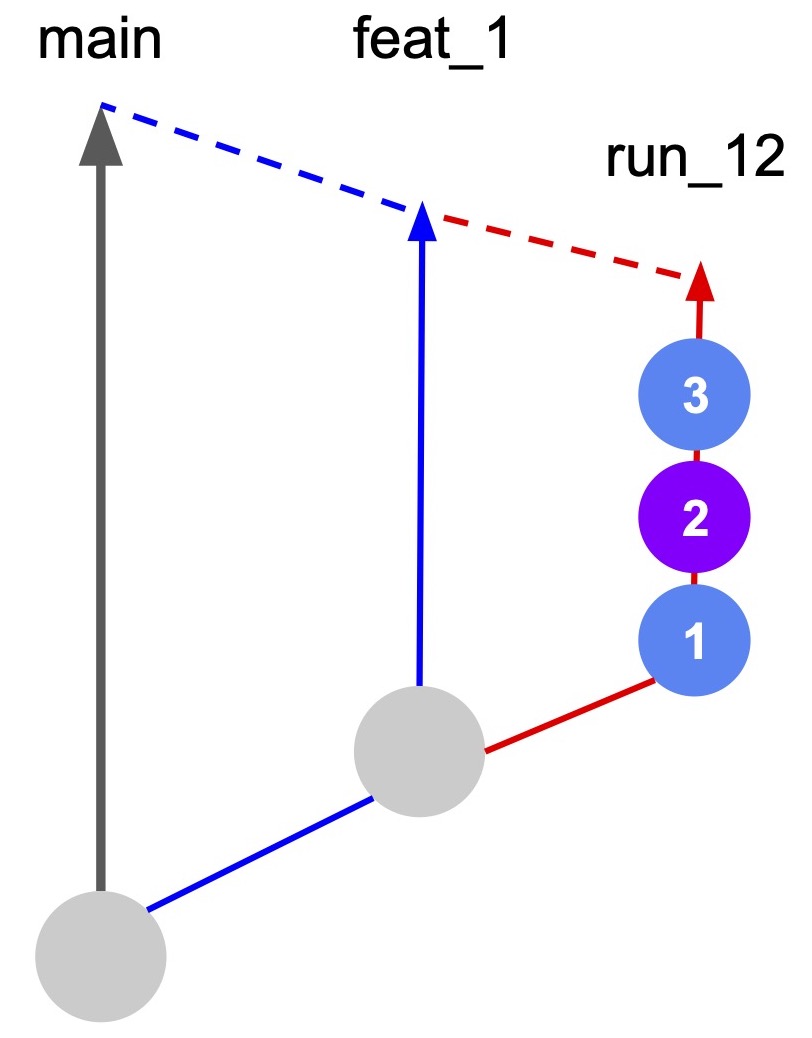}
  \caption{Git semantics for code \textit{and} data. When developing a pipeline, users work on Git branches with associated Nessie branches: every run takes place in an ephemeral branch following the \textit{transform-audit-write} pattern.}
  \label{fig:git}
\end{figure}

Software development best practices and tooling allow developers to work on code (new feature, bug fixing, debugging etc.) in a consistent and sand-boxed way: production code can be cloned, run, modified by developers, but running development code won't leak into a production environment. \texttt{Bauplan} provides the same best practices for data pipelines, enforcing a \textit{transform-audit-write} pattern for all transformations. In particular, we picked \textit{Nessie} \footnote{\url{https://projectnessie.org/}} to provide a git-like semantics: Nessie versions an entire catalog at a time, so it is ideal for transformation use cases when multiple artifacts are affected at each run. Fig.\ref{fig:git} depicts the basic versioning mechanism in the platform:

\begin{enumerate}
    \item the user checkouts through Git a new branch in his project (\textit{feat\_1}), to develop a new pipeline;
    \item in the context of a \texttt{bauplan run} command, \texttt{Bauplan} detects the Git context and creates a Nessie branch with the same name, \textit{feat\_1}, starting from the current production data in the lake \textit{main} branch (grey node); now both the code (through Git) and the data artifacts (through Nessie) are production-like \textit{and} sandboxed;
    \item \texttt{Bauplan} executes the DAG into an ephemeral branch (\textit{run\_12}): by executing each run ``atomically'' we can avoid persisting dirty DAGs -- only when all steps and tests are executed successfully, we are allowed to merge the data into the current branch, making the artifacts \textit{1} and \textit{3} visible to any user with branch access (the obvious analogy here is the concept of \textit{transaction} in databases);
    \item when the merge on \textit{feat\_1} is committed, the ephemeral branch \textit{run\_12} is deleted.
\end{enumerate}

Once again, we remark that we chose to base the developer experience only on Git and the CLI. While we expect users to be familiar with Git, all the data versioning is handled behind the scene transparently. The user is not expected to master Nessie or any of the technologies involved. Instead, they are provided with a sandbox environment for data development with a familiar software-like semantics.

\subsection{Code intelligence}
\label{sec:intelligence}
There is a natural tension in modularity between code and compute: modular code is easier to test, re-use, reason about, on the other side, monolith compute is easier to spin up, manage, orchestrate. The \textit{no infrastructure} principle provides guidance on how to navigate the trade-offs: on the code side, we subscribe to full modularity (e.g. dbt-style transformations), so that each node in the DAG corresponds to one file that is runnable and testable in isolation; on the compute-side, we let the system opting for modularity or monolith depending on the circumstances. In other words, the user is exposed directly only to the \textit{top} layer in Fig.~\ref{fig:plans}: it is the job of the \textit{code intelligence} module (Fig.~\ref{fig:lakehouse}) to take as input the queries and functions defining a pipeline, together with parameters from the CLI, and produce as output first a \textit{logical plan} of operations, and finally a \textit{physical plan} to run the desired transformations.

\subsubsection{From code to the logical plan}
After the pipeline code is ingested (e.g. Section~\ref{app:code}), the full project is snapshotted in an object storage and fingerprinted in a Postgres database, not dissimilarly from what happens for runs in \textit{Metaflow} \cite{Tagliabue2023ReasonableSM}: by assigning an id and immutable artifacts to each run, we guarantee reproducibility for auditing and debugging purposes -- following the \textit{code is data} principle, the same code on the same data version will produce identical results. After versioning, SQL and Python files are parsed: first, logical dependencies are extracted from implicit references -- in our example, \texttt{pickups} is build out of another table (\texttt{SELECT .. FROM trips}), so we need to materialize nodes in the right order; second, environment details for Python functions are extracted -- in our purely functional implementation, a decorator such as \texttt{@requirements} can be used to pin down the needed packages: because of our serverless setup (Section \ref{sec:run}), the OS, container, and environment layers are handled by the system, leaving packages as the only degree of freedom left to control to ensure full reproducibility.

Finally, in our example Python is used only to run an expectation. There is no reasons why Python could not be used to also declare new tables starting from existing ones. In essence, transformations are functional mappers from set of tuples (rows in the ``parent table'') to set of tuples
(rows in the ``child table''): as long as two languages can speak a common dialect over those tuples, they can operate together. 

\subsubsection{The execution plan}
\label{sec:execution}
The output of the parsing step is a logical plan (Fig.~\ref{fig:plans}), so that the system knows which artifacts depends on existing Iceberg tables, which tests need to pass to consider the pipeline healthy, and what needs to be written back into the catalog as a result of running the DAG. The first \texttt{Bauplan} version for executing such a plan was the simplest possible idea, i.e. just mapping the plan to an isomorphic execution, in which each node is executed by one (serverless and stateless) function. However, this naive implementation doesn't optimize around an important feature of data workloads: at RS, computing artifacts is pretty fast, and the bottleneck is often moving data around. To make a concrete example, consider again our sample pipeline: there, the Python expectation is a Pandas function taking a DataFrame as input (the data artifact we are testing), and returning a boolean. Instead of running an Iceberg command first, a SQL query and then a Python function as three separate executions, we pushed down \texttt{WHERE} filters to obtain a smaller in-memory table, then run in-place the SQL logic and the Python expectation. This optimization results in 5x faster feedback loop even with small datasets, and avoid unnecessary spillover to object storage: notably, the user is not required to know any of the underlying implementation details.

\subsection{Serverless runtimes}
\label{sec:run}
When the execution plan is finalized, the computation needs to happen in a fast, reliable, scalable way. Following the functional definitions of pipelines, a serverless runtime is the natural choice in terms of abstraction: the user specifies what needs to happen, the \texttt{Bauplan} platform runs the code in an optimized environment where OS, container, and runtime are under its control \cite{10.5555/3027041.3027047}. In recent years, \textit{serverless} has become an overloaded term, used to vaguely denote a cluster of features not necessarily related \cite{10.1145/3508360,10.1145/3406011} and not necessarily important for (or even, at odds with) data pipelines: scale-to-zero, price-per-second, ``infinite'' and instantaneous concurrency, stateless execution model \cite{10.1145/3360575}.  We identified few essential properties for our serverless platform:

\begin{itemize}
    \item multi-language support with flexible dependencies (Fig.~\ref{fig:lakehouse}): considering SQL code can be run in a Python interpreter connected to object storage (see \textit{duckdb} below), the requirement can be satisfied by a Python runtime allowing an arbitrary combination of interpreter version and dependencies\footnote{Note how the function-first approach provides a level of control -- i.e. specifying packages \textit{per function} -- that is impossible in conventional Spark applications.};
    \item runtime hardware allocation: the same transformation logic should run with 10GB or 20GB of memory depending on the underlying artifacts;
    \item data locality: given that data pipelines are first and foremost about \textit{moving} data, we need to maintain function isolation at the runtime level but allow for shared resources at the artifacts level - moving data is slow and expensive, and object storage should be treated as a last resort \cite{273835};
    \item pausing functions: since a fresh Spark context takes a while to be created, it is typically re-used in a stateful manner. However, since ``freezing'' a container after initialization would make startup time negligible, we could run stateless commands over ephemeral containers.
\end{itemize}

We evaluated AWS Lambda\footnote{\url{https://aws.amazon.com/lambda/}}, 
OpenWhisk\footnote{\url{https://openwhisk.apache.org/}} and OpenLambda\footnote{\url{https://github.com/open-lambda/open-lambda}} as off-the-shelf frameworks, but none of them fully satisfied the \textit{desiderata} above: as typical use cases for serverless are micro-services and glue code in cloud infrastructure, it is not surprising that existing tools would be sub-optimal for our scenarios. Steps in data DAGs have almost opposite requirements when compared to typical functions-as-a-service: startup time is somewhat important, but since the bottleneck is data reading and processing, we play in the 200-1000 ms regime, not 0-200 ms; on the other hand, resources required to compute aggregations require more fine-grained tuning. For these reasons, we invested, as a differentiating feature, in building an orchestration and memory management layer to support workloads in which horizontal scalability is less important than vertical elasticity and efficient data processing.

To support SQL, we leverage \textit{duckdb} \cite{10.1145/3299869.3320212} as our query engine, given its performance, flexibility and full-compatibility with our formats\footnote{An example of running serverless queries has been open-sourced at \url{https://github.com/BauplanLabs/quack-reduce}.}; to support Python, we built custom containerized runtimes and a container manager: furthermore, we were able to exploit the power-law in package utilization \cite{10.5555/3277355.3277362} to limit overall download times with an efficient local, disk-based cache.\footnote{We plan to release a Lambda-based generic runtime for Python functions that leverages object storage for caching.} Our solution allows for fast startup time (300ms), complete runtime isolation at the function level, and customizable sharing policies within the functions in a single DAG execution: as our target deployment model is initially ``Bring Your Own Cloud'', the usual security concerns of multi-tenant virtualization do not apply \cite{246288}.

Finally, we wish to stress that containerization is an active area of research, with exciting possibilities offered by new frameworks such as WASM \cite{RossbergWebAssemblyCoreSpecification}: through an ongoing collaboration with the research group behind SOCK \cite{10.5555/3277355.3277362}, we are actively iterating on this component.

\subsection{Interacting with the platform}
Similar to other popular data tools, interactions between \texttt{Bauplan} users and the platform happen through the CLI, as pipelines get written in the IDE of choice. With the intention of satisfying first the semantics implied by the scenarios in Table \ref{tab:taxonomy}, the CLI experience is centered around two main commands, \textit{query} and \textit{run}:

\begin{itemize}
    \item \texttt{bauplan query -q "SELECT * FROM trips"}: synchronous, point-wise interactions with pre-built artifacts are handled through \textit{query}. As discussed, time-travel is a first-class abstraction, so the same command takes an additional argument to specify the intended branch (if not current): \texttt{-b feat\_1}. 
    \item \texttt{bauplan run}: asynchronous, DAG-long interactions are handled through \textit{run}; starting from the pipeline code in the IDE, issuing \textit{run} starts the intelligence and execution processes depicted in Fig.~\ref{fig:plans}. As DAGs are modular and snapshotted at each execution, additional arguments allow to replay an arbitrary DAG for debugging and inspection: for example, \texttt{-run-id 12 -m pickups+} will re-execute in a sandboxed way the same code over the same data as the run with $id=12$, starting from the \textit{pickups} artifacts and running all its children. 
\end{itemize}

With the goal of truly lowering the bar for data work, the CLI-first approach is easy to learn and easy to extend: in fact, the semantics of \textit{run} mirrors tools that are popular in our user base (\textit{dbt} and \textit{Metaflow}). Moreover, CLI commands are easy for machines to execute as well: since querying and visualizing data in the terminal is not ideal with large datasets, it is trivial to wrap commands in an application layer users are comfortable with, e.g. a dashboard or a Python notebook.

\section{Conclusion and Future work}

We started our journey designing \texttt{Bauplan} by considering -- and dismissing -- two ways to build towards the DLH vision: re-purposing existing Big Data tools, or building a new platform from scratch. Mirroring the Firebolt experience \cite{DBLP:conf/vldb/PasumanskyW22}, we found that re-using existing open source components as initial ``Lego bricks'' can be a powerful third way to getting closer to the goal, without necessarily breaking the bank. While the ``lean startup'' playbook \cite{ries2011startup} of rapid market-driven pivots is not readily applicable to data platforms, re-using components allowed the team to converge quicker to a working end-to-end system, test its strength and weaknesses with early adopters, and place more informed bets on which features are responsible for the greater marginal value.  

There are obviously many other interesting areas that remain to be addressed, e.g. securing data through seamless, yet secure authentication, parallelizing SQL execution, using logs and machine learning to further optimize the experience behind the scenes. Moreover, truly manifesting the DLH vision in the product is a long journey: starting from open source tools was the right choice, but as the platform progresses it is likely we will wander far more into the unknowns to better meet market demands. As Rome was indeed not linted, tested, built nor deployed in a day, we look forward to sharing with the community the next steps of our adventure in future publications.

\begin{acknowledgments}
We are immensely grateful to the open source and data community, and we plan to continue our contributions to open source and open science in this new venture as well. In particular, we wish to thank the PyIceberg, Open Lambda and Nessie teams, with whom we have been collaborating in the past few months while starting \texttt{Bauplan}. Finally, we wish to thank Tyler Caraza-Harter and Ryan Vilim for precious feedback on a previous version of this work.
\end{acknowledgments}

\bibliography{my_bib}

\appendix

\section{Sample data pipeline}

\label{app:code}

We report the full code for the running example of this paper (Section~\ref{sec:example}), as schematically depicted in Fig.~\ref{fig:plans}. Please note that steps are transformed into
a DAG thanks to a simple naming convention: children tables refer to parents (\textit{Step 3} below referring to \textit{Step 1} table), while Python testing functions comply with the \texttt{table\_expectation} syntax.

\textbf{Step 1 (trips)}: read raw data (as stored under an Iceberg table \texttt{taxi\_table}) for a target time window, and extract important columns into a new \texttt{trips} table.

\begin{lstlisting}[language=SQL,label={step1}]
SELECT
    pickup_location_id, 
    passenger_count as count, 
    dropoff_location_id
FROM
    taxi_table
WHERE
    pickup_at >= '2019-04-01'
\end{lstlisting}

\textbf{Step 2 (trips\_expectation)}: we take Step 1 output -- a table named \texttt{trips} --, convert it to a DataFrame and run a statistical check using Python. Similar to declarative data science frameworks such as \textit{Metaflow} \cite{Tagliabue2023ReasonableSM}, Python decorators are used to express directly in code constraints on the target runtime.

\begin{lstlisting}[language=Python,label={step2}]
@requirements({'pandas': '2.0.0'})
def trips_expectation(ctx, trips):
    m = trips['count'].mean()
    return m > 10
\end{lstlisting}

\textbf{Step 3 (pickups)}: we take Step 1 output -- a table named \texttt{trips} --, and produce a new table \texttt{pickups} by aggregating and sorting trip data.

\begin{lstlisting}[language=SQL,label={step3}]
SELECT
    pickup_location_id, 
    dropoff_location_id, 
    COUNT(*) AS counts
FROM
    trips
GROUP BY
    pickup_location_id, 
    dropoff_location_id
ORDER BY
    counts DESC
\end{lstlisting}

\end{document}